\documentstyle[aps,epsfig]{revtex}
\begin{document}

\title{Bogoliubov dispersion relation for a ``photon fluid'': Is this 
a superfluid?\footnote{It is a great pleasure to dedicate this paper 
to my lifelong friend, Marlan Scully.  Marlan and I first met when we 
were graduate students (of Willis Lamb and Charles Townes, 
respectively).  Over many years we have enjoyed discussing and 
learning physics together, as well as sharing and growing in our like 
precious faith.  Happy birthday, Marlan!}} 

\author{Raymond Y. Chiao\\
Dept. of Physics, University of California, Berkeley, 
California 94720-7300\\
E-mail addresses: 
chiao@physics.berkeley.edu}
\date{Version of August 17, 1999}
\maketitle
\begin{abstract}
We discuss the possibility that photons, which are bosons, can form a 
2D superfluid due to Bose-Einstein condensation inside a nonlinear 
Fabry-Perot cavity filled with atoms in their ground states.  A 
``photon fluid'' forms inside the cavity as a result of multiple 
photon-photon collisions mediated by the atoms during a cavity 
ring-down time.  The effective mass and chemical potential for a 
photon inside this fluid are nonvanishing.  This implies the existence 
of a Bogoliubov dispersion relation for the low-lying elementary 
excitations of the photon fluid, and in particular, that sound waves 
exist for long-wavelength, low-frequency disturbances of this fluid.  
Possible experiments to test for the superfluidity of the photon fluid 
based on the Landau critical-velocity criterion will be discussed.\\
\end{abstract}
\section{Introduction}

Inspired by the recent experimental discoveries of Bose-Einstein 
condensation of laser-cooled 
atoms~\cite{Cornell,Hulet,Ketterle,Walls}, we would like to consider 
here the inverse question: Can one observe Bose-Einstein condensation 
of {\em photons}?  Closely related is a second question: Is the 
resulting Bose condensate a superfluid?  We know that photons are 
bosons, so that it would seem that they could in principle undergo 
this kind of condensation.  The difficulty is that in the usual Planck 
blackbody configuration, which consists of an empty 3D cavity, the 
photon is massless and its chemical potential is zero, so that the 
Bose-Einstein condensation of photons under these circumstances would 
seem to be impossible.

However, we consider here an atom-filled 2D Fabry-Perot cavity 
configuration, instead of the usual empty 3D Planck cavity 
configuration.  We find that if one excites only one of the 
longitudinal modes of the Fabry-Perot cavity by means of 
narrow-linewidth laser radiation, so that the dynamics of the light 
inside the cavity becomes effectively two-dimensional~\cite{2D}, and 
if this radiation is well detuned to the red side of the resonance of 
the atoms in their ground state, so that an effective repulsive 
photon-photon interaction mediated by the atoms results, the resulting 
effective mass and chemical potential of a photon inside the cavity no 
longer vanishes.  Thus Bose-Einstein condensation of photons inside 
the Fabry-Perot cavity {\em can} occur.  We shall explore the 
circumstances under which this may happen, and shall connect this 
problem with an earlier problem solved by Bogoliubov for the 
weakly-interacting Bose gas.  In this way, we shall see that the 
Bogoliubov dispersion relation should hold for the ``photon fluid'' 
that forms as a result of multiple photon-photon collisions occurring 
inside the cavity.  In particular, we shall see that sound waves, or 
``phonons'' in the photon fluid, are the lowest-lying excitations of 
the system.  According to an argument due to Landau, this implies that 
the photon fluid could become a superfluid.

Historically speaking, in the study of the interaction of light with 
matter, most of the recent emphasis has been on exploring new states 
of matter, such as the recently observed atomic Bose-Einstein 
condensates.  However, not as much attention has been focused on 
exploring new states of light.  Of course, the invention of the laser 
led to the discovery of a new state of light, namely the coherent 
state, which is a very robust one.  Two decades ago, squeezed states 
were discovered, but these states are not as robust as the coherent 
state, since they are easily degraded by scattering and absorption.  
In contrast to the laser, which involves a population-inverted atomic 
system that is far away from thermal equilibrium, we shall explore 
here states very close to the ground state of a photonic system, and 
hence very near absolute zero in temperature.  Hence they should be 
robust ones.
 
 The interacting many-photon system (the ``photon fluid'') is an 
 example of a quantum many-body problem.  In an extension of the 
 interacting Bose gas problem, we shall derive the Bogoliubov 
 dispersion relation for the weakly-interacting photon gas with 
 repulsive photon-photon interactions, starting from the microscopic 
 (i.e., the second-quantized) level.  Thereby we shall find an 
 expression for the effective chemical potential of a photon in the 
 photon fluid, and shall relate the velocity of sound in the photon 
 fluid to this nonvanishing chemical potential.  In this way, we shall 
 lay the theoretical foundations for an experiment to measure the 
 sound wave part of the dispersion relation for the photon fluid.  We 
 shall also propose another experiment to measure the critical 
 velocity of this fluid, and thus to test for the possibility of the 
 superfluidity of the resulting state of the light.
 
 Although the interaction Hamiltonian used in this paper is equivalent 
 to that used earlier in four-wave squeezing, we emphasize here the 
 {\em collective} aspects of the problem which result from multiple 
 photon-photon collisions.  This leads to the idea of the ``photon 
 fluid.''  It turns out that microscopic and macroscopic analyses 
 yield exactly the same Bogoliubov dispersion relation for the 
 excitations of this fluid~\cite{PRA}.  Hence it may be argued that 
 there is nothing fundamentally new in the microscopic analysis given 
 below which is not already contained in the macroscopic, classical 
 nonlinear optical analysis.  However, it is the microscopic analysis 
 which leads to the viewpoint of the interacting photon system as a 
 ``photon fluid,'' a conception which could give rise to new ways of 
 understanding and discovering nonlinear optical phenomena.  
 Furthermore, the interesting question of the quantum optical state of 
 the light inside the cavity resulting from multiple collisions 
 between the photons (i.e., whether it results in a coherent, 
 squeezed, Fock, or some other quantum state), cannot be answered by 
 the macroscopic, classical nonlinear optical analysis, and this 
 necessitates the microscopic treatment given below.
 
\section{The Bogoliubov problem}

Here we re-examine one particular many-body problem, the one first 
solved by Bogoliubov~\cite{Bogoliubov=1947,Pines}.  Suppose that one 
has a zero-temperature system of bosons which are interacting with 
each other repulsively, for example, a dilute system of small, bosonic 
hard spheres.  Such a model was intended to describe superfluid 
helium, but in fact it did not work well there, since the interactions 
between atoms in superfluid helium were too strong for the theory for 
be valid.  In order to make the problem tractable theoretically, let 
us assume that these interactions are weak.  In the case of light, the 
interactions between the photons are in fact always weak, so that this 
assumption is a good one.  However, these interactions are 
nonvanishing, as demonstrated by the fact that photon-photon 
collisions mediated by atoms excited near, but off, resonance have 
been experimentally observed~\cite{Morgan}.  We start with the 
Bogoliubov Hamiltonian
\begin{eqnarray}
H &=& H_{free}+H_{int}\nonumber\\
H_{free}&=&\sum_{p} \epsilon(p)a_{p}^{\dagger}a_{p}\nonumber\\
H_{int}&=&\frac{1}{2}\sum_{\kappa  pq} 
 V(\kappa  ) a_{p+\kappa  }^{\dagger} a_{q-\kappa  }^{\dagger}a_{p}a_{q}\,,
 \label{1}
\end{eqnarray}
where the operators $a_{p}^{\dagger}$ and $a_{p}$ are 
creation and annihilation operators, respectively, for bosons with 
momentum $p$, which satisfy the Bose commutation relations
\begin{equation}
[a_{p},a_{q}^{\dagger}] = \delta_{pq}\mbox{ and } [a_{p},a_{q}] = 
[a_{p}^{\dagger},a_{q}^{\dagger}] = 0\,.
\end{equation}

The first term $H_{free}$ in the Hamiltonian represents the energy of 
the free boson system, and the second term $H_{int}$ represents the 
energy of the interactions between the bosons arising from the 
potential energy $V(\kappa )$, which is the Fourier transform of the 
potential energy $V(r_{2}-r_{1})$ in configuration space of a pair of 
bosons located at $r_{2}$ and $r_{1}$.  The interaction term is 
equivalent to the one responsible for producing squeezed states of 
light via four-wave mixing~\cite{Slusher}.  It represents the 
annihilation of two particles, here photons, of momenta $p$ and $q$, 
along with the creation of two particles with momenta $p+\kappa $ and 
$q-\kappa $, in other words, a scattering process with a momentum 
transfer $\kappa $ between a pair of particles with initial momenta 
$p$ and $q$, along with the assignment of an energy $V(\kappa )$ to 
this scattering process.  Here the assumption that the interactions 
are weak means that the second term in the Hamiltonian is much smaller 
than the first, i.e, $|V(\kappa )|\ll|\epsilon(\kappa )|$.

\section{The free-photon dispersion relation inside a Fabry-Perot 
resonator}

\begin{figure}
\centerline{\psfig{figure=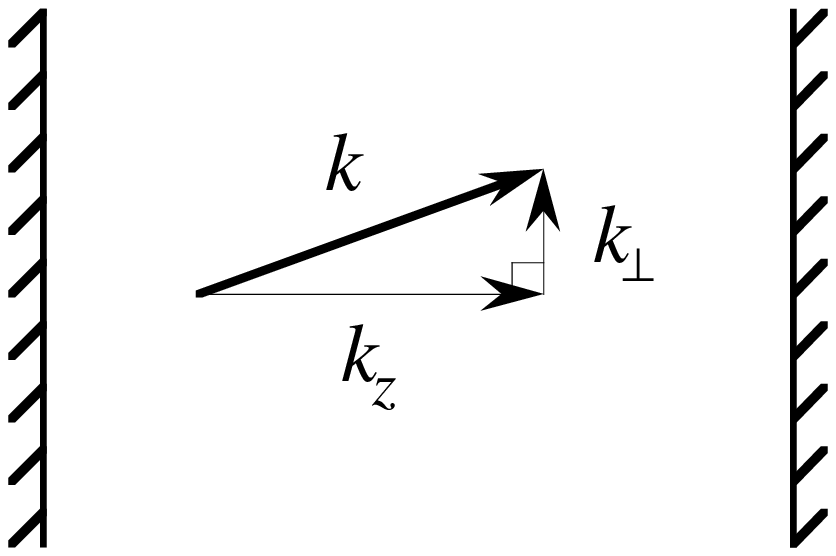,width=7cm}}
\caption{A planar Fabry-Perot imposes boundary conditions which 
quantize the allowed values of $k_{z}$, where $z$ is the axis normal 
to the mirrors, in units of $\pi/L$, where $L$ is the separation of 
the mirrors.  For a plane-wave mode which propagates at a small angle 
with respect to the $z$ axis, there arises an effective 
nonrelativistic energy-momentum relation for an noninteracting, 
trapped 2D photon, whose effective mass is $m \approx 
\hbar\omega/c^{2}$ (see text).}
\label{Fabry-Perot}
\end{figure}

Photons with momenta $p$ and $q$ also obey the above commutations 
relations, so that the Bogoliubov theory should in principle also 
apply to the weakly-interacting photon gas.  The factor $\epsilon(p)$ 
represents the energy as a function of the momentum (the dispersion 
relation) for the free, i.e., noninteracting, bosons.  In the case of 
photons in a Fabry-Perot resonator, the boundary conditions of the 
mirrors cause the $\epsilon(p)$ of a photon trapped inside the 
resonator to become an energy-momentum relation which is identical to 
that of a nonrelativistic particle with an effective 
mass~\cite{Morgan,Garrison} of $m=\hbar\omega/c^{2}$.  This can be 
understood starting from Fig.~\ref{Fabry-Perot}.
 
For high-reflectivity mirrors, the vanishing of the electric field at 
the reflecting surfaces of the mirrors imposes a quantization 
condition on the allowed values of the $z$-component of the photon 
wave vector, $k_{z}=n\pi/L$, where $n$ is an integer, and $L$ is the 
distance between the mirrors.  Thus the usual frequency-wavevector 
relation
\begin{equation}
\omega(k) = c[k_{x}^{2}+k_{y}^{2}+k_{z}^{2}]^{1/2}\,,
\end{equation}
upon multiplication by $\hbar$, becomes the energy-momentum 
relation for the photon 
\begin{equation}
E(p) = c[p_{x}^{2}+p_{y}^{2}+p_{z}^{2}]^{1/2} = 
c[p_{x}^{2}+p_{y}^{2}+\hbar^{2}n^{2}\pi^{2}/L^{2}]^{1/2}
=c[p_{x}^{2}+p_{y}^{2}+m^{2}c^{2}]^{1/2}\,,
\end{equation}
where $m={\hbar}n\pi/Lc$ is the effective mass of the 
photon.  In the limit of small-angle (or paraxial) propagation, where 
the small transverse momentum of the photon satisfies the inequality
\begin{equation}
p_{\perp}= [p_{x}^{2}+p_{y}^{2}]^{1/2}\ll p_{z}=\hbar k_{z} = \hbar 
n\pi/L\,,
\end{equation}
we obtain from a Taylor expansion of the relativistic relation, a 
nonrelativistic energy-momentum relation for the 2D noninteracting 
photons inside the Fabry-Perot resonator
\begin{equation}
E(p_{\perp}) \cong mc^{2}+p_{\perp}^{2}/2m\,,
\label{nonrelativistic}
\end{equation}
where $m={\hbar}n\pi/Lc\cong \hbar\omega/c^{2}$ is the effective mass 
of the confined photons.  It is convenient to redefine the zero of 
energy, so that only the effective kinetic energy,
\begin{equation}
\epsilon(p_{\perp}) \cong p_{\perp}^{2}/2m\,,
\label{7}
\end{equation}
remains.  To establish the connection with the Bogoliubov Hamiltonian, 
we identify the two-dimensional momentum $p_{\perp}$ as the momentum 
$p$ that appears in this Hamiltonian, and the above $\epsilon(p_{\perp})$ 
as the $\epsilon(p)$ that appears in Eq.~(\ref{1}).

\section{The Bogoliubov dispersion relation for the photon fluid}

Now we know that in an ideal Bose gas at absolute zero temperature, 
there exists a Bose condensate consisting of a macroscopic number 
$N_{0}$ of particles occupying the zero-momentum state.  This feature 
should survive in the case of the weakly-interacting Bose gas, since 
as the interaction vanishes, one should recover the Bose condensate 
state.  Hence following Bogoliubov, we shall assume here that even in 
the presence of interactions, $N_{0}$ will remain a macroscopic number 
in the photon fluid\cite{Thouless}.  This macroscopic number will be 
determined by the intensity of the incident laser beam which excites 
the Fabry-Perot cavity system, and turns out to be a very large number 
compared to unity (see below).  For the ground state wave function 
$\Psi_{0}(N_{0})$ with $N_{0}$ particles in the Bose condensate in the 
$p=0$ state, the zero-momentum operators $a_{0}$ and $a_{0}^{\dagger}$ 
operating on the ground state obey the relations
\begin{eqnarray}
a_{0}\left|\Psi_{0}(N_{0})\right>&=&
\sqrt{N_{0}}\left|\Psi_{0}(N_{0}-1)\right>\nonumber \\
a_{0}^{\dagger}\left|\Psi_{0}(N_{0})\right>&=&
\sqrt{N_{0}+1}\left|\Psi_{0}(N_{0}+1)\right>\,.
\end{eqnarray}
Since $N_{0}\gg1$, we shall neglect the difference between the factors 
$\sqrt{N_{0}+1}$ and $\sqrt{N_{0}}$.  
Thus one can 
replace all occurrences of $a_{0}$ and $a_{0}^{\dagger}$ by the 
$c$-number $\sqrt{N_{0}}$, so that to a good approximation 
$[a_{0},a_{0}^{\dagger}]\approx 0$.  
However, the number of particles in 
the system is then no longer exactly conserved, as can be seen by 
examination of the term in the Hamiltonian
\begin{equation}
\sum_{\kappa }V(\kappa )a_{\kappa }^{\dagger}a_{-\kappa 
}^{\dagger}a_{0}a_{0}\approx N_{0}\sum_{\kappa }V(\kappa )a_{\kappa 
}^{\dagger}a_{-\kappa }^{\dagger}\,,
\end{equation}
which represents the creation of a pair of particles, i.e., photons, with 
transverse momenta $\kappa  $ and $-\kappa  $ out of nothing.  

\begin{figure}
\centerline{\psfig{figure=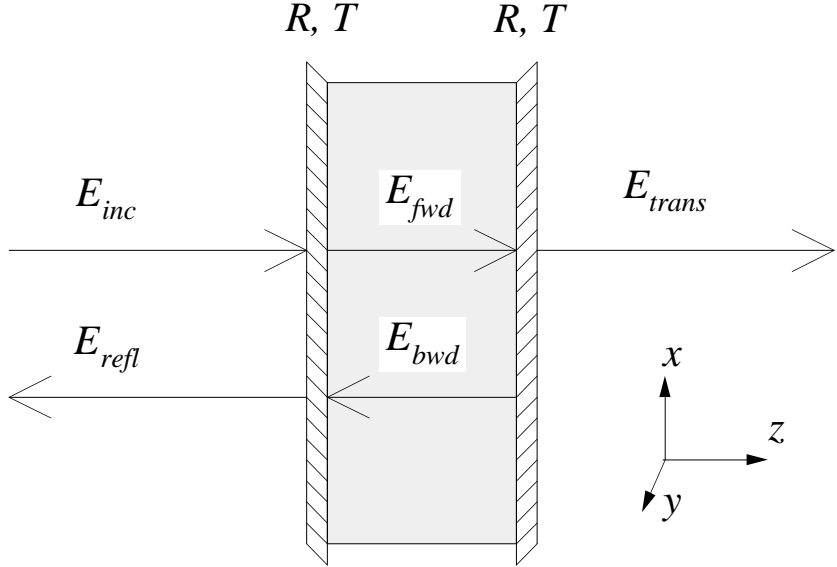,width=11cm}}
\caption{Fields and coordinate system in the Fabry-Perot cavity.  The 
applied field $E_{inc}$ arises from a laser beam incident from the 
left.  An atomic vapor excited to the red side of resonance by the 
incident light fills the space (the gray area) between the two 
mirrors.  The presence of these atoms leads to a self-defocussing Kerr 
nonlinearity (corresponding to repulsive photon-photon interactions) 
inside the cavity.}
\label{cavityfig}
\end{figure}

However, whenever the system is open one, i.e., whenever it is 
connected to an external reservoir of particles which allows the total 
particle number number inside the system (i.e., the cavity) to 
fluctuate around some constant average value, then the total number of 
particles need only be conserved on the average.  Formally, one 
standard way to compensate for the lack of exact particle number 
conservation is to use the Lagrange multiplier method and subtract a 
chemical potential term $\mu N_{op}$ from the Hamiltonian (just as in 
statistical mechanics when one goes from the canonical ensemble to the 
grand canonical ensemble)~\cite{Hugenholtz}
\begin{equation}
H \rightarrow H'= H - \mu N_{op},
\label{grand}
\end{equation}
where $N_{op}=\sum_{p}a_{p}^{\dagger}a_{p}$ is the total number 
operator, and $\mu$ represents the chemical potential, i.e., the 
average energy for adding a particle to the open system described by 
$H$.  In the present context, we are considering the case of a 
Fabry-Perot cavity with low, but finite, transmittivity mirrors which 
allow photons to enter and leave the cavity, due to an input light 
beam coming in from the left and an output beam leaving from the right 
(see Fig.~\ref{cavityfig}).  This permits a realistic physical 
implementation of the external reservoir, since the Fabry-Perot cavity 
allows the total particle number inside the cavity to fluctuate due to 
particle exchange with the beams outside the cavity.  However, the 
photons remain trapped inside the cavity long enough so that a 
condition of thermal equilibrium is achieved after multiple 
photon-photon interactions (i.e., after very many collisions, which is 
indeed the case for the experimental numbers to be discussed below).  
This leads to the formation of a photon fluid inside the 
cavity~\cite{mu}.

It will be useful to separate out the zero-momentum components of the 
interaction Hamiltonian, since it will turn out that there is a 
macroscopic occupation of the zero-momentum state due to 
Bose condensation.  The prime on the sums
$\sum'_{p}$, $\sum'_{p\kappa}$, and $\sum'_{\kappa pq}$ in the
following equation denotes sums over momenta explicitly
excluding the zero-momentum state, i.e., all the running indices
$p$, $\kappa$, $q$,$p+\kappa$,$q-\kappa$ which are not explicitly
set equal to zero, are nonzero:
\begin{eqnarray}
H_{int}&=&\frac{1}{2}V(0)a_{0}^{\dagger} 
a_{0}^{\dagger}a_{0}a_{0} + 
V(0){\sum_{p}}'a_{p}^{\dagger}a_{p}a_{0}^{\dagger}a_{0}+ \nonumber\\
&&{\sum_{p}}'\left(V(p)a_{p}^{\dagger}a_{0}^{\dagger}a_{p}a_{0} 
+\frac{1}{2}\left[V(p)a_{p}^{\dagger}a_{-p}^{\dagger}a_{0}a_{0} + 
V(p)a_{0}^{\dagger}a_{0}^{\dagger}a_{p}a_{-p}\right]\right)+\nonumber\\
&&{\sum_{p\kappa }}'V(\kappa ) \left(a_{p+\kappa 
}^{\dagger}a_{0}^{\dagger}a_{p}a_{\kappa } +a_{p+\kappa 
}^{\dagger}a_{-\kappa }^{\dagger}a_{p}a_{0}\right) 
+\frac{1}{2}{\sum_{\kappa pq}}'V(\kappa ) \left(a_{p+\kappa 
}^{\dagger} a_{q-\kappa }^{\dagger}a_{p}a_{q}\right)\,.
\label{10}
\end{eqnarray}
Here we have also assumed that $V(p)=V(-p)$.  By thus separating out 
the zero-momentum state from the sums in the Hamiltonian, and 
replacing all occurrences of $a_{0}$ and $a_{0}^{\dagger}$ by 
$\sqrt{N_{0}}$, we find that the Hamiltonian $H'$ in Eq.~(\ref{grand}) 
decomposes into three parts
\begin{equation}
H'=H_{0} + H_{1} + H_{2}\,,
\end{equation}
where, in decreasing powers of $\sqrt N_{0}$,
\begin{equation}
H_{0} = \frac{1}{2}V(0)a_{0}^{\dagger} 
a_{0}^{\dagger}a_{0}a_{0} \approx \frac{1}{2}V(0)N_{0}^{2}\,, 
\label{H0}
\end{equation}
\begin{equation}
H_{1}\approx {\sum_{p}}'\epsilon'(p)a_{p}^{\dagger}a_{p} 
+\frac{1}{2}N_{0}{\sum_{p}}'V(p)\left(a_{-p}^{\dagger}a_{p}^{\dagger} 
+a_{-p}a_{p}\right)\,,
\label{original}
\end{equation}
\begin{equation}
H_{2}\approx \sqrt{N_{0}}\;{\sum_{p\kappa }}'V(\kappa  )
\left(a_{p+\kappa  }^{\dagger}a_{p}a_{\kappa  }
+a_{p+\kappa  }^{\dagger}a_{-\kappa  }^{\dagger}a_{p}\right)
+\frac{1}{2}{\sum_{\kappa  pq}}'V(\kappa  )
\left(a_{p+\kappa  }^{\dagger} a_{q-\kappa  }^{\dagger}a_{p}a_{q}\right)\,.
\end{equation}
Here
\begin{equation}
\epsilon'(p) = \epsilon(p) + N_{0}V(0) + N_{0}V(p) - \mu
\label{primed}
\end{equation}
is a modified photon energy, where $N_{0}$ is given by
\begin{equation}
N_{0} + <\Psi_{0}| {\sum_{p}}'a_{p}^{\dagger}a_{p}
|\Psi_{0}> = N\,
\end{equation}
(the term $<\Psi_{0}| {\sum_{p}}'a_{p}^{\dagger}a_{p} |\Psi_{0}>$ 
represents the number of photons in the ``depletion,'' i.e., those 
photons which have been scattered out of the condensate at $T=0$ due 
to photon-photon collisions), and where $\mu$ is given by
\begin{equation}
\mu =\frac{\partial E_{0}}{\partial N}\,.
\end{equation}
Here $E_{0}=\left<\Psi_{0}|H|\Psi_{0}\right>$ is the ground state 
energy of $H$.  In the approximation that there is little depletion of 
the Bose condensate due to the interactions (i.e., $N\approx 
N_{0}\gg1$), the first term of Eq.~(\ref{10}) (i.e., $H_{0}$ in 
Eq.~(\ref{H0})) dominates, so that
\begin{equation}
E_{0}\approx \frac{1}{2}N_{0}^{2}V(0)
\approx\frac{1}{2}N^{2}V(0),
\end{equation}
and therefore that
\begin{equation}
\mu\approx NV(0)\approx N_{0}V(0).
\label{chemical}
\end{equation}
This implies that the effective chemical potential of a photon, i.e., 
the energy for adding a photon to the photon fluid, is given by the 
number of photons in the Bose condensate times the repulsive pairwise 
interaction energy between photons with zero relative momentum.  It 
should be remarked that the fact that the chemical potential is 
nonvanishing here makes the thermodynamics of this two-dimensional 
photon system quite different from the usual three-dimensional, Planck 
blackbody photon system.  It should also be remarked that the 
conventional wisdom which tells us that Bose-Einstein condensation and 
superfluidity are impossible in 2D bosonic systems, does not apply 
here.  To the contrary, we believe that superfluidity of the 
topological, 2D Kosterlitz-Thouless kind (with algebraic decay of long 
range order) is possible for the photon fluid~\cite{KT}.

In the same approximation $N\approx N_{0}\gg1$, Eq.~(\ref{primed}) 
becomes, upon use of the fact that $\mu\approx N_{0}V(0)$
\begin{equation}
\epsilon'(p) \approx \epsilon(p) + N_{0}V(p).
\label{prime}
\end{equation}
This is the single-particle photon energy in the Hartree 
approximation.  Here it is again assumed that $|H_{1}|\gg|H_{2}|$, 
i.e., that the interactions between the bosons are sufficiently weak, 
so as not to deplete the Bose condensate significantly.  In the case 
of the weakly-interacting photon gas inside the Fabry-Perot resonator, 
since the interactions between the photons are indeed weak, this 
assumption is a good one.

Following Bogoliubov, we now introduce the following canonical 
transformation in order to diagonalize the quadratic-form 
Hamiltonian $H_{1}$ in Eq.~(\ref{original}):
\begin{eqnarray}
\alpha_{\kappa  }&=&u_{\kappa  }a_{\kappa  }+v_{\kappa  }a_{-\kappa  }^{\dagger}\nonumber \\
\alpha_{\kappa }^{\dagger}&=&u_{\kappa }a_{\kappa 
}^{\dagger}+v_{\kappa }a_{-\kappa  }\,.
\label{canonical}
\end{eqnarray}
Here $u_{\kappa  }$ and $v_{\kappa  }$ are two real $c$-numbers 
which must satisfy the condition
\begin{equation}
u_{\kappa  }^{2}-v_{\kappa  }^{2}=1\,,
\label{18}
\end{equation}
in order to insure that the Bose commutation relations are preserved 
for the new creation and annihilation operators for certain new 
quasi-particles, $\alpha_{\kappa  }^{\dagger}$ and $\alpha_{\kappa  }$, i.e., that
\begin{equation}
[\alpha_{\kappa  },\alpha_{\kappa  '}^{\dagger}]=\delta_{\kappa  ,\kappa  '}
\mbox{ and }[\alpha_{\kappa  },\alpha_{\kappa  '}]= 
[\alpha_{\kappa  }^{\dagger},\alpha_{\kappa  '}^{\dagger}]=0\,.
\end{equation}
We seek a diagonal form of $H_{1}$ given by
\begin{equation}
H_{1}={\sum_{\kappa }}'\left[\tilde{\omega} (\kappa )\left(\alpha_{\kappa 
}^{\dagger}\alpha_{\kappa } 
+\frac{1}{2}\right)+\mbox{constant}\right]\,,
\label{diagonal}
\end{equation}
where $\tilde{\omega}(\kappa )$ represents the energy of a 
quasi-particle of momentum $\kappa $.  Substituting the new 
creation and annihilation operators $\alpha_{\kappa }^{\dagger}$ and 
$\alpha_{\kappa }$ given by Eq.~(\ref{canonical}) into 
Eq.~(\ref{diagonal}), and comparing with the original form of the 
Hamiltonian $H_{1}$ in Eq.~(\ref{original}), we arrive at the 
following necessary conditions for diagonalization:
\begin{eqnarray}
\tilde{\omega} (\kappa  )u_{\kappa  }v_{\kappa  }&=&\frac{1}{2}N_{0}V(\kappa  ) \label{21}\\
u_{\kappa  }^{2}&=& \frac{1}{2}\left[1+\epsilon'(\kappa  )/\tilde{\omega} (\kappa  )\right] \label{22}\\
v_{\kappa  }^{2}&=&\frac{1}{2}\left[-1+\epsilon'(\kappa  )/\tilde{\omega} (\kappa  )\right]. \label{23}
\end{eqnarray}
Squaring Eq.~(\ref{21}) and substituting from 
Eqs.~(\ref{22}) and (\ref{23}), we obtain
\begin{equation}
\tilde{\omega} (\kappa  )^{2}=\epsilon'(\kappa  )^{2}-N_{0}^{2}V(\kappa  )^{2}
\approx \epsilon(\kappa  )^{2}+ 2\epsilon(\kappa  )N_{0}V(\kappa  ),
\label{square}
\end{equation}
where in the last step we have used Eq.~(\ref{prime}).

\begin{figure}
\centerline{\psfig{figure=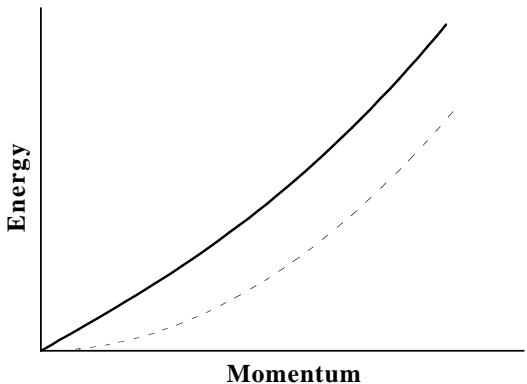,width=7cm}}
\caption{The energy versus momentum of an elementary excitation in the 
weakly-interacting Bose gas, here in the present case, the photon 
fluid.  The solid line represents the Bogoliubov dispersion relation 
given by Eq.~(\ref{28}), for the special case that $V(\kappa  ) \approx
V(0) =$ constant, and the dashed line represents a quadratic 
dispersion relation for a noninteracting, diffracting photon inside 
the Fabry-Perot resonator.}
\label{Bogoliubov}
\end{figure}

Thus the final result is that the Hamiltonian $H_{1}$ in 
Eq.~(\ref{diagonal}) describes a collection of {\em noninteracting} 
simple harmonic oscillators, i.e., quasi-particles, or elementary 
excitations of the photon fluid from its ground state.  The 
energy-momentum relation of these quasi-particles is obtained from 
Eq.~(\ref{square}) upon substitution of $\epsilon(\kappa )=\kappa 
^{2}/2m$ from Eq.~(\ref{7})
\begin{equation}
\tilde{\omega} (\kappa ) \approx \left[\frac{\kappa ^{2} N_{0} 
V(\kappa )}{m} + \frac{\kappa ^{4}}{4m^{2}}\right]^{1/2},
\label{28}
\end{equation}
which we shall call the ``Bogoliubov dispersion relation.''  This 
dispersion relation is plotted in Fig.~\ref{Bogoliubov}, in the 
special case that $V(\kappa ) \approx V(0) =$ constant.  (Note that 
Landau's roton minimum can also be incorporated into this theory by a 
suitable choice of the functional form of $V(\kappa )$.)

For small values of $\kappa $ this dispersion relation is {\em linear} 
in $\kappa $, indicating that the nature of the elementary excitations 
here is that of {\em phonons}, which in the classical limit of large 
phonon number leads to sound waves propagating inside the photon 
fluid at the sound speed
\begin{equation}
v_{s}= \lim_{\kappa \rightarrow 0}\frac{\tilde{\omega}(\kappa 
)}{\kappa} = 
\left(\frac{N_{0}V(0)}{m}\right)^{1/2}=\left(\frac{\mu}{m}\right)^{1/2}\,.
\label{29}
\end{equation}
At a transition momentum $\kappa  _{c}$ given by
\begin{equation}
\kappa  _{c}= 2\left(mN_{0}V(\kappa  _{c})\right)^{1/2}
\end{equation}
(i.e., when the two terms of Eq.~(\ref{28}) are 
equal), the linear relation between energy and momentum turns into a 
quadratic one, indicating that the quasi-particles at large momenta 
behave essentially like nonrelativistic free particles with an energy 
of $\kappa  ^{2}/2m$.  The reciprocal of $\kappa  _{c}$ defines a characteristic 
length scale
\begin{equation}
\lambda_{c}\equiv 2\pi\hbar /\kappa  _{c}=\pi\hbar /mv_{s}\,,
\end{equation}
which characterizes the distance scale over which collective effects 
arising from the pairwise interaction between the photons become 
important.

Thus in the above analysis, we have shown that all the approximations 
involved in the Bogoliubov theory should be valid ones for the case of 
the 2D photon fluid inside a nonlinear Fabry-Perot cavity.  Hence the 
Bogoliubov dispersion relation should indeed apply to this fluid; in 
particular, there should exist sound wave modes of propagation in the 
photon fluid.  As additional evidence for the existence of these 
modes, we have recently found that the same Bogoliubov dispersion 
relation emerges from a classical nonlinear optical analysis of this 
problem~\cite{PRA}, which we shall not reproduce here.  The velocity 
of sound found by the macroscopic, classical nonlinear optical 
analysis is identical to the one found in Eq.~(\ref{29}) for the 
velocity of phonons in the photon fluid in the above microscopic 
analysis, provided that one identifies the energy density of the light 
inside the cavity with the number of photons in the Bose condensate as 
follows:
\begin{equation}
{\cal E}_{0}^{2} = 8\pi N_{0}\hbar \omega/{V}_{cav}\;,
\label{40}
\end{equation}
where ${V}_{cav}$, the cavity volume, is also the quantization volume 
for the electromagnetic field, and provided that one makes use of the 
known proportionality between the Kerr coefficient $n_{2}$ and the 
photon-photon interaction potential $V(0)$~\cite{Deutsch,Akhmanov}
\begin{equation}
V(0) = 8 \pi (\hbar \omega)^{2}n_{2}/V_{cav}\;.
\end{equation}

In fact, the entire dispersion relation found by the classical, 
macroscopic analysis for sound waves associated with fluctuations in 
the light intensity inside a resonator filled with a self-defocusing 
Kerr medium, turns out to be formally identical to the above 
Bogoliubov dispersion relation obtained quantum mechanically for the 
elementary excitations of the photon fluid, in the approximation 
$V(\kappa ) \approx V(0) =$ constant.  This is a valid approximation, 
since the pairwise interaction potential between two photons is given 
by a transverse 2D pairwise spatial Dirac delta function, whose 
strength is proportional to $n_{2}$, provided that the photons 
propagate paraxially, and the nonlinearity is fast.  It should be kept 
in mind that the phenomena of self-focusing and self-defocusing in 
nonlinear optics can be viewed as arising from \textit{pairwise 
interactions} between photons when the light propagation is paraxial 
and the Kerr nonlinearity is fast~\cite{Deutsch,Akhmanov}.  Since in a 
quantum description the light inside the resonator is composed of 
photons, and since these photons as the constituent particles are 
weakly interacting repulsively with each other through the 
self-defocusing Kerr nonlinearity to form a photon fluid, this formal 
identification between the microscopic and macroscopic results for the 
Bogoliubov relation is a natural one~\cite{correspondence}.

\begin{figure}
\centerline{\psfig{figure=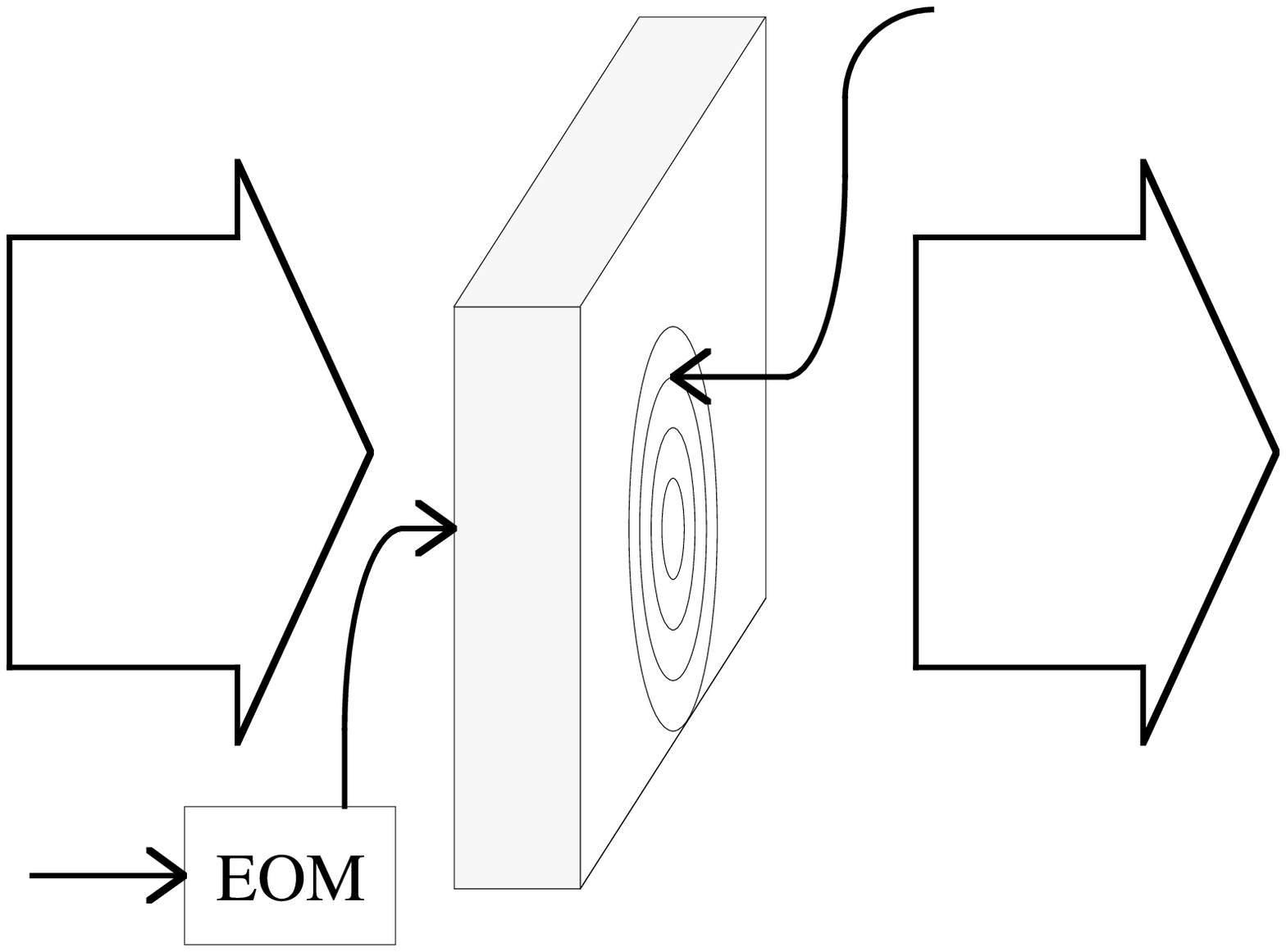,width=7cm}}
\caption{Schematic of an experiment to observe the sound waves in a 
photon fluid which fills a nonlinear Fabry-Perot resonator.  The 
nonlinear medium (denoted by the gray area) consists of atoms excited 
by a broad laser beam (denoted by the broad incoming arrow) well 
detuned to the red side of resonance.  The goal is the verify the 
Bogoliubov dispersion relation, Eq.~(\ref{28}).  An electro-optic 
modulator (EOM) modulates the intensity of light at a radio frequency, 
which is then injected by means of an optical fiber tip at a single 
point on the entrance face of the Fabry-Perot resonator.  The 
wavelength of the resulting sound waves in the photon fluid can be 
measured by scanning in the transverse direction the tip of another 
optical fiber across the output face of the Fabry-Perot.  A $2\pi$ 
phase shift of the modulated pick-up signal relative to that of the 
EOM modulation signal corresponds to a transverse displacement of the 
tip by a sound wavelength.}
\label{experimentfigure}
\end{figure}

One possible experiment to see these sound waves is sketched in 
Fig.~\ref{experimentfigure}.  The sound wave mode is most simply 
observed by applying two incident optical fields to the nonlinear 
cavity: a broad plane wave resonant with the cavity to form the 
nonlinear background fluid on top of which the sound waves can 
propagate, and a weaker amplitude-modulated beam which is modulated at 
the sound frequency in the radio range by an electro-optic modulator, 
and injected by means of an optical fiber tip at a single point on the 
entrance face of the Fabry-Perot.  The resulting weak time-varying 
perturbations in the background light induce transversely propagating 
density waves in the photon fluid, which propagate away from the point 
of injection like ripples on a pond.  This sound wave can be 
phase-sensitively detected by another fiber tip placed at the exit 
face of the Fabry-Perot some transverse distance away from the 
injection point, and its sound wavelength can be measured by scanning 
this fiber tip transversely across the exit face.

The experiment could employ a cavity length $L$ of $2$ cm and mirrors 
with reflectivities of $R=0.997$ for a cavity finesse ${\cal F}=1050$.  
The optical nonlinearity could be provided by rubidium vapor at 
$80^{\rm o}$ C, corresponding to a number density of $10^{12}$ 
rubidium atoms per cubic centimeter.  Incident on the cavity could be 
a circularly-polarized CW laser beam, detuned by around $600$ MHz to 
the red side of a closed two-level transition, for example, the 
$\left|F=2, m_{F}=+2\right>\,\rightarrow\,\left|F'=3, 
m_{F'}=+3\right>$ transition of the $^{87}\rm Rb$ $D_{2}$ line.  Thus 
the Kerr nonlinear coefficient could be that of a pure two-level 
atomic system virtually excited well off resonance (i.e., with a 
detuning much larger than the absorption linewidth), which was 
calculated by Grischkowsky~\cite{Grischkowsky}:
\begin{equation}
n_{2}=\pi N_{atom\,} \mu^{4}/\hbar^{3}\Delta^{3}\,\approx \, 6\times 
10^{-6}\,\rm{cm^{3}/erg}\,\approx \, 5\times 
10^{-8}\,\rm{cm^{2}/Watt},
\end{equation}
where $N_{atom}$ is the atomic number density of the atomic vapor, 
$\mu$ is the matrix element of the two-level atomic system, and 
$\Delta$ is the detuning of the laser frequency from the atomic 
resonance frequency.  Thus the $\Delta\,\approx \,$ 600 MHz detuning 
of the laser from the atomic resonance used in the above example would 
be considerably larger than the Doppler width of 340 MHz of the 
rubidium vapor, and the residual absorption arising from the tails of 
the nearby resonance line would give rise to a loss which would be 
less than or comparable to the loss arising from the mirror 
transmissions.  This extra absorption loss would contribute to a 
slightly larger effective cavity loss coefficient, but would not 
otherwise alter the qualitative behavior of the Bogoliubov dispersion 
relation.  The conditions of validity for the microscopic Bogoliubov 
theory should be well satisfied by these experimental parameters.  An 
intracavity intensity of $40\,\mathrm{W/cm^{2}}$ would result in 
$\Delta n=|n_{2}|{{\cal E}_{0}}^{2}\,\approx\,2\times10^{-6}$, for a 
sound speed $v_{s}\,\approx\,4\times10^{7}\,\mathrm{cm/s}$.  For this 
intensity, $N_{0} \,\approx\,8\times 10^{11}$, so that the condition 
for the validity of the Bogoliubov theory $N_{0}\gg1$ should be well 
satisfied.  The cavity ring-down time $\tau_{cav}=2{\cal 
F}L/c\,\approx\,0.14\,\mu\rm s$ would be much longer than the mean 
photon-photon collision time $\tau_{coll}=(12\omega n_{2}|{\cal 
E}_{0}|^{2})^{-1}\,\approx\, 17\,\rm{ps}$, so that a photon fluid 
should indeed form inside the cavity, since there would be 
approximately 8000 photon-photon collisions within a cavity ring-down 
time, so that the assumption of thermal equilibrium should be a valid 
one.

It should be noted that the above Bogoliubov theory is not limited to 
the above two-level atomic Kerr nonlinearity, which was chosen only 
for the purposes of illustration.  One could replace this two-level 
nonlinearity with other recent, more promising kinds of 
nonlinearities, such as that in a four-level system, where absorption 
could be eliminated by the use of quantum interference while the Kerr 
nonlinearity could be simultaneously enhanced~\cite{Imamoglu}, or such 
as that due to photon exchange, where the nonlinearity is proportional 
to $N_{atom}^{2}$ rather than to $N_{atom}$~\cite{Franson}.

\section{Discussion}

We suggest here that the Bogoliubov form of dispersion relation, 
Eq.~(\ref{28}), implies that the photon fluid formed by the repulsive 
photon-photon interactions in the nonlinear cavity is actually a 
photon {\em superfluid}.  This means that a superfluid state of light 
might actually exist.  Although the exact definition of superfluidity 
is presently still under discussion, especially in light of the 
question whether the recently discovered atomic Bose-Einstein 
condensates are superfluids or not~\cite{Walls}, one indication of the 
existence of a photon superfluid would be that there exists a critical 
transition from a dissipationless state of superflow, i.e., a laminar 
flow of the photon fluid below a certain critical velocity past an 
obstacle, into a turbulent state of flow, accompanied by energy 
dissipation associated with the shedding of a von-Karman street of 
$quantized$ vortices past this obstacle, above this critical velocity.  
(It is the generation of {\em quantized} vortices above this critical 
velocity which distinguishes the onset of {\em superfluid} turbulence 
from the onset of {\em normal} hydrodynamic turbulence.)

The physical meaning of the Bogoliubov dispersion relation is that the 
lowest energy excitations of the system consist of quantized sound 
waves or phonon excitations in a superfluid, whose maximum critical 
velocity is then given by the sound wave velocity.  By inspection of 
this dispersion relation, a single quantum of any elementary 
excitation cannot exist with a velocity below that of the sound wave.  
Hence no excitation of the superfluid is possible at all for any 
object moving with a velocity slower than that of the sound wave 
velocity, according to an argument by Landau~\cite {Landau}.  Hence 
the flow of the superfluid must be dissipationless below this critical 
velocity.  Above a certain critical velocity, dissipation due to 
vortex shedding is expected from computer simulations based on the 
Gross-Pitaevskii (or Ginzburg-Landau or nonlinear Schr\"{o}dinger) 
equation, which should give an accurate description of this system at 
the macroscopic level~\cite{Pomeau}.

We propose a follow-up experiment to demonstrate that the sound wave 
velocity, typically a few thousandths of the vacuum speed of light, is 
indeed a maximum critical velocity of a fluid, i.e., that this photon 
fluid exhibits persistent currents in accordance with the Landau 
argument based on the Bogoliubov dispersion relation.  Suppose we 
shine light at some nonvanishing incidence angle on a Fabry-Perot 
resonator (i.e., exciting it on some off-axis mode).  This light 
produces a uniform flow field of the photon fluid, which flows inside 
the resonator in some transverse direction and at a speed determined 
by the incidence angle.  A cylindrical obstacle placed inside the 
resonator will induce a laminar flow of the superfluid around the 
cylinder, as long as the flow velocity remains below a certain 
critical velocity.  However, above this critical velocity a turbulent 
flow will be induced, with the formation of a von-Karman vortex street 
associated with quantized vortices shed from the boundary of the 
cylinder~\cite{Pomeau}.  The typical vortex core size is given by the 
light wavelength divided by the square root of the nonlinear index 
change.  Typically the vortex core size should be around a few hundred 
microns, so that this nonlinear optical phenomenon should be readily 
observable.

A possible application is suggested by an analogy with the Meissner 
effect in superconductors, or the Hess-Fairbank effect in superfluid 
helium: Vortices in an incident light beam would be expelled from the 
interior of the photon superfluid.  This would lead to a useful 
beam-cleanup effect, in which speckles in a dirty incident laser beam 
would be expelled upon transmission through the nonlinear Fabry-Perot 
resonator, so that a clean, speckle-free beam emerges.

\section*{Acknowledgments} I thank Jack Boyce for performing the 
classical calculation, and for making the first attempt to do the 
sound wave experiment, and L.M.A. Bettencourt, D.A.R. Dalvit, I.H. 
Deutsch, J.C. Garrison, D. H. Kobe, D.H. Lee, M.W. Mitchell, J. 
Perez-Torres, D.S. Rokhsar, D.J. Thouless, E.M. Wright, and W.H. Zurek 
for helpful discussions.  The work was supported by the ONR and by the 
NSF.

\pagebreak

\section*{FIGURES}

\begin{thebibliography}{99}

\bibitem{Cornell} 
M. H. Anderson, J. R. Ensher, M. R. Matthews, C. E .Wiemann, and E. A. 
Cornell, Science {\bf 269}, 198 (1995).

\bibitem{Hulet} 
C. C. Bradley, C. A. Sackett, J. J. Tollett, and R. G. Hulet, 
Phys.~Rev.~Lett.~{\bf 75}, 1687 (1995).

\bibitem{Ketterle} 
K. B. Davis, M.-O. Mewes, M. R. Andrews, N. J. van Druten, D. S. 
Durfee, D. M. Kurn, and W. Ketterle, Phys.~Rev.~Lett.~{\bf 75}, 3969 
(1995).

\bibitem{Walls} 
For a review of recent work on dilute-gas Bose condensates, see A. S. 
Parkins and D. F. Walls, Physics Reports {\bf 303}, 1 (1998).

\bibitem{2D}
The dynamics of the light inside the cavity becomes effectively 
two-dimensional, if the longitudinal mode spacing of the Fabry-Perot is 
much larger than the laser linewidth, so that only a single 
longitudinal mode is excited by the incident laser beam.

\bibitem{PRA}
R. Y. Chiao and J. Boyce, Phys.~Rev.~A (to be published; LANL 
e-print quant-ph/9905001).

\bibitem{Bogoliubov=1947} 
N. Bogoliubov, J. Phys.~(U.S.S.R.) {\bf 11}, 23 (1947).

\bibitem{Pines} 
D. Pines, {\it The Many-Body Problem} (Benjamin, New York, 1961).

\bibitem{Morgan} 
M. W. Mitchell (private communication).

\bibitem{Garrison}
I. H. Deutsch and J. C. Garrison, Phys.~Rev.~A {\bf 43}, 2498 (1991).

\bibitem{Thouless}
Since we have assumed a zero-temperature Bose gas, following 
Bogoliubov we start this calculation with the ground state of the 
system in the macroscopically occupied zero-momentum Fock or number 
state $\left|N_{0},p=0\right>$.  However, it is also possible to 
derive the same Bogoliubov dispersion relation starting from a system 
in a coherent state $\left|\alpha,p=0\right>$, where $|\alpha|>>1$.  I 
thank Prof.~David Thouless for sharing with me his unpublished notes 
concerning this last point.

\bibitem{Slusher} 
R. E. Slusher {\it et al.}, J. Opt.~ Soc.~ Am.~B {\bf 4}, 1453 (1987).

\bibitem{Hugenholtz} 
N. M. Hugenholtz and D. Pines, Phys.~Rev.~{\bf 116}, 489 (1959); G. 
W. Goble and D. H. Kobe, Phys.~Rev.~A {\bf 10}, 851 (1974).  

\bibitem{mu}
Note that the quantum mechanics problem we solve here is the 
$T\rightarrow 0$ limit of the more general statistical mechanics 
problem at $T\neq 0$, so that the chemical potential method (i.e., the 
Lagrange multiplier method) used in Eq.~(\ref{grand}) should also be 
valid in the purely quantum mechanical limiting case.  Note also that 
the results obtained here are independent of the details of the 
reservoir and of its coupling to the photon fluid system inside the 
cavity, when there exists thermodynamic equilibrium.  The photon fluid 
is in thermal equilibrium with the external light beams (see 
Fig.~\ref{cavityfig}) in the same sense that a superfluid helium thin 
film on a cold substrate is in thermal equilibrium with the helium 
atoms incident on it from the vapor, and also with the helium atoms 
evaporating from it into the vapor.

\bibitem{KT}
A universal, two-dimensional topological phase transition of the 
Kosterlitz-Thouless type should also be possible in the 2D photon 
fluid.  Due to the existence of topological excitations, i.e., 
vortices, in this fluid, the well-known theorems which deny the 
possibility of Bose-Einstein condensation and of superfluidity (i.e., 
of true off-diagonal long-range order) in two-dimensional systems do 
not apply.


\bibitem{Deutsch}
I. H. Deutsch, R. Y. Chiao, J. C. Garrison, Phys.~Rev.~Lett.~{\bf 69}, 
3627 (1992).

\bibitem{Akhmanov}
R. Y. Chiao, I. H. Deutsch, J. C. Garrison, and E. W. Wright, in {\it 
Frontiers in Nonlinear Optics: the Serge Akhmanov Memorial Volume}, H. 
Walther, N. Koroteev, and M. O. Scully, eds., Institute of Physics 
Publishing, Bristol and Philadelphia, 1993, p.  151-182.

\bibitem{correspondence}
One may wonder why the classical nonlinear optical calculation gives 
exactly the same result as the quantum many-body calculation.  An 
answer is that one expects classical sound waves to have the same 
dispersion relation as phonons in a quantum many-body system: there 
exists a classical, correspondence-principle limit of the quantum 
many-body problem, in which the collective excitations (i.e., their 
dispersion relation) do not change their form in the classical limit 
of large phonon number.

\bibitem{Grischkowsky}
D. Grischkowsky, Phys.~Rev.~Lett.~{\bf 24}, 866 (1970).

\bibitem{Imamoglu}
A. Imamoglu, H. Schmidt, G. Woods, and M. Deutsch, 
Phys.~Rev.~Lett.~{\bf 79}, 1467 (1997).

\bibitem{Franson}
J. D. Franson, Phys.~Rev.~Lett.~{\bf 78}, 3852 (1997); also to be 
published in Phys. Rev. A.

\bibitem{Landau}
L. D. Landau and E. M. Lifshitz, {\it Statistical Physics} (Pergamon, 
London, 1958), p. 202.

\bibitem{Pomeau} 
Y. Pomeau and S. Rica, Comptes Rendus de l'Acad\'{e}mie des Sciences 
(Paris) {\bf 317}, 1287 (1993).

\end{thebibliography}
\end{document}